\documentclass[12pt]{article}
\usepackage{authblk}
\usepackage{graphicx}
\usepackage{amssymb}
\usepackage{amsmath}
\title{
Analog gravity in nonisentropic fluids}

\author[]{Neven Bili\'c\thanks{bilic@irb.hr}\hspace{2pt} and Hrvoje Nikoli\'c\thanks{hnikolic@irb.hr}}
\affil[]{Division of Theoretical Physics, Rudjer Bo\v skovi\'c Institute, 10002 Zagreb, Croatia}

\date{\today}
\begin{document}
\maketitle
\begin{abstract}

The analog acoustic metric has been originally derived for adiabatic acoustic perturbations propagating in 
an isentropic irrotational ideal fluid.
In the framework of a Lagrangian hydrodynamic description we demonstrate that 
under certain conditions the usual acoustic metric can be derived 
for nonisentropic fluids. 
In a special case when  the pressure
takes a special form and the nonadiabatic perturbations are neglected  
the adiabatic acoustic perturbations corresponding to massless phonons
propagate in an analog metric of the usual type. 
\end{abstract}

%


\section{Introduction}
The possibility that a pseudo-Riemannian geometry of spacetime
can be mimicked by fluid dynamics in Minkowski spacetime has been exploited in various contexts
including 
emergent gravity \cite{babichev,novello2}, scalar theory of gravity \cite{novello}, 
and acoustic geometry \cite{visser,bilic,kinoshita, barcelo2}. 
The basic idea is the emergence of an effective metric of the form
\begin{equation}
G_{\mu\nu} = a [g_{\mu\nu}-(1-c_{\rm s}^2)u_\mu u_\nu],
\label{eq100}
\end{equation}
which describes the effective geometry for acoustic perturbations propagating in 
a fluid potential flow with  $u_\mu\propto \partial_\mu \theta$.
The quantity $c_{\rm s}$ is the adiabatic speed of sound, 
the conformal factor $a$ is 
related to the equation of state of the fluid, 
and the background spacetime metric $g_{\mu\nu}$ is usually assumed  Minkowski. 
In an equivalent field-theoretical picture the fluid velocity $u_\mu$ is derived from the scalar field as 
$u_\mu= \partial_\mu \theta/\sqrt{X}$ and $a$ and $c_{\rm s}$ are expressed in terms of the Lagrangian and its first 
and second derivatives with respect to the kinetic energy term $X=g^{\mu\nu}\theta_{,\mu}\theta_{,\nu}$.
The effective metric (\ref{eq100}) has been originally derived 
for an isentropic irrotational perfect fluid. However, it has been recently demonstrated that the condition 
of vanishing vorticity
can be relaxed for a Bose Einstein condensate coupled to the electromagnetic field
\cite{cropp}.

In a slightly different context,
the metric of the form (\ref{eq100}) has been used to show that a pseudo-Riemann 
spacetime with Lorentz signature may be derived from a Riemann metric with Euclidean signature 
\cite{barbero1,barbero2, mukohyama}. In that case, the vector $u_\mu$ represents the normalized
gradient of a hypothetical scalar field which governs the dynamics and the signature of the effective
spacetime.


In applications of analog geometry, in addition 
to the energy-momentum conservation or the Euler equation, the continuity equation is usually assumed. 
However, with this assumption some interesting geometries cannot be mimicked by analog geometry.
For example,  the Schwarzschild metric cannot be mimicked by the 
non-relativistic version of (\ref{eq100}) 
unless the continuity equation is abandoned  \cite{barcelo2,visser2} and the same holds true in the relativistic case.
 The fluid in which the particle number is not conserved is generally nonisentropic.
Therefore, it is of considerable interest to study analog geometries in nonisentropic fluids.

The propagation of sound in an ideal isentropic fluid is automatically  an adiabatic process.
In a nonisentropic fluid flow the propagation of perturbations can be adiabatic (i.e., at fixed entropy)
or nonadiabatic. In a field theoretical description Babichev, Mukhanov and Vikman have shown \cite{babichev} that
the scalar-field perturbations propagate in the  analog metric
of the form (\ref{eq100}) and acquire an effective mass.
In their approach the perturbations are generally nonadiabatic.

The work presented here is partially motivated  by a recent article of S.~Hossenfelder
\cite{hossenfelder}. She has derived an analog metric that mimics the
geometry of a planar black hole (BH) in asymptotic Anti de Sitter (AdS) space. 
A careful analysis which will be  provided in section \ref{planar} of the present paper
reveals that the fluid in this model is essentially nonisentropic.
In the present paper we will study adiabatic and nonadiabatic perturbations 
propagating in a general nonisentropic fluid. It turns out that 
under certain conditions, if nonadiabatic perturbations are ignored, 
the usual analog gravity description applies with
the effective metric in the usual form (\ref{eq100}). In this case
the phonons corresponding to adiabatic acoustic perturbations remain massless.  

We divide the remainder of the paper into four sections and an appendix.
We start with section  \ref{nonisentropic}, in which  
we give a hydrodynamic and field-theoretic  description of a nonisentropic fluid.
In the following section, section \ref{acoustic}, we study  perturbations of the flow
and derive conditions under which the equation of motion can be written in the form
of a Klein-Gordon equation in an analog curved spacetime.
In section \ref{planar} we apply our formalism to the model of
an analog planar AdS$_5$ BH.
Concluding remarks are given in section \ref{conclude}.
Finally, in appendix \ref{second} we provide a brief account of the second law of thermodynamics
relevant for nonisentropic fluid flows.

\section{Nonisentropic flow}
\label{nonisentropic}


 The analog acoustic geometry has been derived under strict requirements of energy momentum conservations,
 particle number conservation,  and 
  vanishing vorticity. 
The first two restrictions are sufficient conditions for adiabaticity.
If a stronger restriction of isentropy is assumed together with vanishing vorticity, 
the velocity field may be expressed as $w u_\mu =\theta_{,\mu}$
where $\theta$ is the velocity potential and $w$ is the specific enthalpy. 
The reverse of the above statement is not true: a potential flow alone implies
only vanishing vorticity and 
implies neither particle number conservation nor isentropy.

Next we  consider a nonisentropic fluid flow
and derive general properties of relativistic nonisentropic fluids. 
To make it clear which properties are consequences of which assumptions, 
each subsection studies consequences of one additional assumption, 
from more fundamental assumptions towards less fundamental ones.

\subsection{Energy-momentum conservation}

The most fundamental  assumption is the energy-momentum conservation 
\begin{equation}
 {T^{\mu\nu}}_{;\nu}=0.
  \label{eq402}
 \end{equation}
The energy-momentum tensor of an ideal relativistic fluid 
can be expressed as
\begin{equation}
T_{\mu\nu}=(p+\rho) u_{\mu}u_{\nu}-p g_{\mu\nu},
\label{eq019}
\end{equation}
where $p$ and $\rho$ are the fluid pressure and energy density, respectively,
and $g_{\mu\nu}$ is the background metric with
signature $(+---)$. 
The contraction of (\ref{eq402}) with $u_{\mu}$ 
gives 
\begin{equation}
u^\mu\rho_{,\mu}+(p+\rho){u^\mu}_{;\mu}=0.
 \label{eq441}
\end{equation}
Inserting this into (\ref{eq402}) with (\ref{eq019}) gives the relativistic Euler equation
\cite{landau}
\begin{equation}
(p+\rho)u^{\nu}u_{\mu;\nu}
-p_{,\mu}
+u^{\nu} p_{,\nu}u_{\mu}  =0.
\label{eq003}
\end{equation}

\subsection{The first law of thermodynamics}
For a general 
thermodynamic system at nonzero temperature $T$ the first law of thermodynamics 
may be written as
\begin{equation}\label{rnif5}
dp=ndw -nTds ,
\end{equation}
where $n$ is the particle number density,  $s$ is the specific entropy, i.e., the entropy per particle, 
and $w$ is the specific enthalpy defined as
\begin{equation}
w=\frac{p+\rho}{n} .
 \label{eq404}
\end{equation}
Since the pressure can be viewed as a function of two variables
$s$ and $w$, a comparison of  
the total derivative 
\begin{equation}
dp=\frac{\partial p}{\partial w} dw+ \frac{\partial p}{\partial s}ds ,
\label{eq417}
\end{equation}
with (\ref{rnif5})
yields the thermodynamic relations
\begin{equation}
n=\left. \frac{\partial p}{\partial w}\right|_s, 
\quad  Tn= \left.-\frac{\partial p}{\partial s}\right|_w.
 \label{eq418}
\end{equation}
The second equation, which may be understood as a defining equation for the temperature,
shows that in a realistic system the pressure is a 
non-increasing function of specific entropy at fixed specific enthalpy.

Using (\ref{eq404}) and expressing (\ref{rnif5}) as 
$p_{,\mu}=nw_{,\mu}-nTs_{,\mu},$ from
Eq.~(\ref{eq441}) it follows
\begin{equation}\label{rnif7}
w(nu^{\mu})_{;\mu}+nTu^{\mu}s_{,\mu}=0 .
\end{equation}
Similarly, Eq.~(\ref{eq003}) with (\ref{rnif5}) gives
\begin{equation}\label{rnif8}
u^{\nu}(wu_{\mu})_{;\nu}-w_{,\mu}=
T(u^{\nu}s_{,\nu}u_{\mu}-s_{,\mu}) .
\end{equation}
The sign of $(nu^{\mu})_{;\mu}$ determines whether we have 
local particle creation or destruction: the particles are locally created or destroyed
if $(nu^{\mu})_{;\mu}$ is positive or negative, respectively.
Hence, assuming that $w$ is positive, 
Eq.~(\ref{rnif7}) states that the entropy per particle 
increases when the number of particles decreases. 

In the following considerations the temperature will play no role.
Therefore,  we will use Eq. (\ref{rnif7}) in the form
\begin{equation}
(nu^{\mu})_{;\mu}=\frac{1}{w}\frac{\partial p}{\partial s}u^{\mu}s_{,\mu} .
\label{eq442}
\end{equation}
Cearly, the particle number is generally not conserved.
If the particle number were conserved, 
i.e. if the continuity equation $(nu^{\mu})_{;\mu}=0$ were true, 
Eq.~(\ref{eq442}) would imply the adiabatic condition $u^{\mu}s_{,\mu}=0$. 
It is worth mentioning that in most applications of thermodynamics and fluid dynamics in cosmology
a conservation of particle number and entropy has been assumed (see, e.g., \cite{saridakis} and references therein).

\subsection{Potential flow}

Some of the equations will further simplify if we assume that the enthalpy flow 
$wu_{\mu}$ is a gradient of a scalar potential, i.e., if there exist a scalar function $\theta$ such that
the velocity field satisfies \cite{landau}
\begin{equation}
w u_\mu =\partial_\mu\theta .
 \label{eq403}
\end{equation}
In this case 
the left-hand side of (\ref{rnif8}) vanishes identically,
so Eq.~(\ref{rnif8}) reduces to
\begin{equation}\label{rnif10}
s_{,\mu}=u^{\nu}s_{,\nu}u_{\mu} .
\end{equation}
Hence, in a potential flow the entropy gradient is proportional to the gradient of the potential.
The assumption (\ref{eq403}) is automatically satisfied in the field-theory formalism, which will be discussed next.

\subsection{Field-theory formalism}
\label{field}

In order to give a more precise meaning to the quantity 
$u^\mu s_{,\mu}$ which, generally, may be an arbitrary function of $w$ and $s$, 
it proves convenient  to use the field-theoretical description of
 fluid dynamics \cite{garriga,bilic4}. 
  Consider a Lagrangian ${\cal L}(X,\theta)$
 that depends 
 on a dimensionless scalar field  $\theta$ 
 and on the kinetic energy term 
 \begin{equation}
X = g^{\mu \nu} {\theta}_{, \mu}
\theta_{, \nu}.
\end{equation}
The corresponding energy-momentum tensor is given by
\begin{equation}
T_{\mu\nu}= 2{\cal L}_X
\theta_{,\mu}\theta_{,\nu}
-{\cal L}g_{\mu\nu} ,
\label{eq508}
\end{equation}
where the subscript $X$ 
denotes a partial derivative with respect to $X$.
For $X>0$, the energy-momentum tensor  
takes the perfect fluid
form (\ref{eq019})
where  the quantities  
\begin{equation}
p ={\cal L} , \quad
\rho = 2 X {\cal L}_{X}-{\cal L} , 
\label{eq4003}
\end{equation}
and
\begin{equation}
 u_\mu=\frac{\partial_\mu \theta}{\sqrt{X}}
 \label{eq4012}
\end{equation}
are
 the  pressure, 
energy density, and velocity of the fluid, respectively.
Obviously, the field $\theta$ serves as the velocity potential and
comparing (\ref{eq4012}) with (\ref{eq403}) we identify the specific enthalpy as
\begin{equation}
w=\sqrt{X}.
 \label{eq409}
\end{equation}
From (\ref{eq403}) and (\ref{eq4003}) we find the
expression for the particle number density 
\begin{equation}
 n=2\sqrt{X}{\cal L}_X.
 \label{eq2112}
\end{equation}
It is understood that the quantities defined in equations (\ref{eq4003})-(\ref{eq2112}) are 
derived from an on-shell Lagrangian, i.e., from  the Lagrangian in which the field 
$\theta$ is a solution
to the  equation of motion
\begin{equation}
(2 {\cal L}_X \, g^{\mu\nu} \theta_{,\mu})_{;\nu}
-\partial{\cal L}/\partial\theta
=0,
\label{eq400}
\end{equation}
which may be written as 
\begin{equation}
(n u^\mu)_{;\mu}
=\partial\mathcal{L}/\partial\theta .
\label{eq401}
\end{equation}
If the Lagrangian were a function of $X$ only, the right-hand side of (\ref{eq401}) would vanish
and this equation would expresses a conservation of the current $J_\mu=n u_\mu$.
A comparison of (\ref{eq401}) and (\ref{rnif7}) demonstrates that the conservation of $J_\mu$ is closely related  
to the isentropy of the fluid flow. 

If the right-hand side of (\ref{eq401}) were zero,
i.e., if $\partial{\mathcal{L}}/\partial\theta=0$, equation (\ref{rnif5}) and  the definitions (\ref{eq4003})-(\ref{eq2112})
would imply $ds=0$.  
In other words, if $\mathcal{L}$ is a function of $X$ only the fluid flow is necessarily isentropic.
The reverse is also true, although in somewhat weaker sense \cite{piattella}: if $ds=0$ 
there exists a field redefinition $\tilde{\theta}=\tilde{\theta}(\theta)$ such that 
the function $\mathcal{L}=\mathcal{L}(X,\theta)$ can be brought to the form $\mathcal{L}=\mathcal{L}(\tilde{X})$,
where $\tilde{X}=g^{\mu\nu}\tilde{\theta}_\mu\tilde{\theta}_\nu$.

Next, we demonstrate that there exist a functional relationship between the specific entropy $s$
and $\theta$.
Using the definitions (\ref{eq409}) and (\ref{eq2112}) from (\ref{eq4003}) it follows
\begin{equation}
dp=\mathcal{L}_X dX+\frac{\partial\mathcal{L}}{\partial\theta}d\theta=
n dw +\frac{\partial\mathcal{L}}{\partial\theta} d\theta .
 \label{eq411}
\end{equation}
Comparing this with (\ref{rnif5}) we conclude that keeping $\theta$ fixed is equivalent to
keeping the specific entropy $s$ fixed and hence 
\begin{equation}
\left.\frac{\partial p}{\partial w}\right|_{s}=\left. \frac{\partial p}{\partial w}\right|_\theta=n .
 \label{eq412}
\end{equation}
Moreover, equations (\ref{rnif5}), (\ref{eq411}), and (\ref{eq412}) demonstrate that the the specific entropy $s$
is a function of the  field $\theta$  such that
\begin{equation}
\frac{ds}{d\theta}
=\frac{\partial\mathcal{L}/\partial \theta}{\partial p/\partial s} .
 \label{eq415}
\end{equation}
Then, a comparison between (\ref{eq401}) and (\ref{eq442}) yields
\begin{equation}
u^\mu s_{,\mu}= w f(s),
 \label{eq437}
\end{equation}
where 
\begin{equation}
f(s)=\frac{ds}{d\theta} .
 \label{eq438}
\end{equation}
Clearly,
the functional relationship $s=s(\theta)$ depends on the Lagrangian.
For example, if the Lagrangian is a function of the kinetic term $X$ only, i.e., if
$\partial\mathcal{L}/\partial \theta=0$, the right-hand side of (\ref{eq415}) will vanish 
yielding $s={\rm const}$, i.e., an isentropic fluid.
It has been argued \cite{piattella} that in most cases one can  identify $\theta$ with $s$.
However this identification cannot be generally correct since, as we have noted above,
the flow  could be isentropic even if $\mathcal{L}(X,\theta)$ were a nontrivial function of $\theta$.

Hence, motivated by the Lagrangian description of fluid dynamics,
in the following considerations
 we will use 
Eq.\ (\ref{eq442}) in the form  
\begin{equation}
(n u^\mu)_{;\mu}
=f(s)\frac{\partial p}{\partial s}
\label{eq439}
\end{equation}
without specifying the function $f(s)$.
This is the key equation which will be used in the next section to derive a
propagation  equation for linear perturbations.


\section{Acoustic metric}
\label{acoustic}

Acoustic metric is the effective metric 
 perceived by acoustic perturbations propagating in a perfect fluid
background. 
Under certain conditions
the perturbations satisfy a Klein-Gordon equation in curved geometry 
with metric of the form (\ref{eq100}).

We first derive a propagation equation for linear perturbations 
of a nonisentropic flow assuming a fixed background geometry.
Given  some average bulk motion represented by
 $p$, $n$, and
 $u^{\mu}$, following the standard procedure \cite{visser,bilic,landau}, 
we make a replacement
 \begin{equation}
p\rightarrow p+\delta p, \quad n\rightarrow n+\delta n ,
\quad
u^{\mu}\rightarrow u^{\mu}+\delta u^{\mu},
\label{eq008}
\end{equation}
where the small disturbances 
$\delta p$, 
$\delta n$, and 
$\delta u^{\mu}$ 
are induced by the perturbations $\delta w$ and $\delta s$
of two independent variables $w$ and $s$:
 \begin{equation}
w\rightarrow w+\delta w, 
\quad
s\rightarrow s+\delta s.
\label{eq007}
\end{equation}
Then
equation
(\ref{eq439})
at linear order yields
\begin{equation}
\left[\left(\frac{\partial n}{\partial w}\delta w +\frac{\partial n}{\partial s}\delta s \right)u^\mu
+n \delta u^\mu\right]_{;\mu}=
f \frac{\partial n}{\partial s}\delta w+
\frac{\partial}{\partial s} \left( f \frac{\partial p}{\partial s}\right) \delta s,
 \label{eq408}
\end{equation}
where we have employed (\ref{eq412}) in the first term on the right-hand side.
In contrast to the previous works, in this equation  
we have  nonadiabatic terms  related to the perturbation $\delta s$,
in addition to the adiabatic terms related to $\delta w$. 
From 
(\ref{eq403}) 
it follows
\begin{equation}
\delta w=u^\mu\delta\theta_{,\mu},
 \label{eq406}
\end{equation}
\begin{equation}
w\delta u^\mu=(g^{\mu\nu}-u^\mu u^\nu)\delta\theta_{,\nu}.
 \label{eq407}
\end{equation}
To simplify the notation, in the following we introduce a perturbation $\sigma$
such that $\delta s=f \sigma$ and denote by $\chi$ the perturbation
$\delta\theta \equiv \chi$ appearing in the variations $\delta w$ and $\delta u^\mu$.
Hence,  $\chi$ and $\sigma$ represent  adiabatic 
and nonadiabatic perturbations, respectively.
Then, combined with (\ref{eq406}) and (\ref{eq407}), equation (\ref{eq408}) takes the form
\begin{equation}
\left(f^{\mu\nu}
\chi_{,\nu} \right)_{;\mu}
-f\frac{\partial n}{\partial s}u^\mu(\chi_{,\mu}-\sigma_{,\mu})
+\left[\left( f
\frac{\partial n}{\partial s}u^\mu\right)_{;\mu}
-f\frac{\partial}{\partial s}\left(f\frac{\partial p}{\partial s}\right)\right]\sigma=0,
 \label{eq413}
\end{equation}
where 
\begin{equation}
f^{\mu\nu}=\frac{n}{w}\left[g^{\mu\nu} -\left(1-\frac{w}{n}\frac{\partial n}{\partial w}\right)u^\mu u^\nu\right] .
 \label{eq423}
\end{equation}

Next we  derive conditions under which Eq. (\ref{eq413})
can be written as the Klein-Gordon equation for the acoustic perturbation $\chi$ 
in an effective curved geometry. 
Obviously, equation (\ref{eq413}) with (\ref{eq423})  
in its most general form  cannot be written
as a Klein-Gordon equation 
because the term 
\begin{equation}
f \frac{\partial n}{\partial s}u^\mu\chi_{,\mu}
 \label{eq419}
\end{equation}
introduces an extra coupling between the velocity field and the derivative of $\chi$. 
However, if we impose certain restrictions on perturbations,
the term (\ref{eq419}) could be eliminated. In that regard we distinguish two cases:
a) nonadiabatic perturbations with $\sigma\equiv \chi$ and b) purely adiabatic perturbations with $\sigma=0$.

\subsection{Nonadiabatic perturbations with $\sigma\equiv \chi$} 
In the 
field theoretical context,  as in, e.g.,  Ref.\ \cite{babichev}
it seems natural to identify $\sigma\equiv \chi$.
This is because in the variation of the Lagrangian  one does not distinguish   
the perturbation
$\delta\theta$ of the field $\theta$
in the explicit functions of $\theta$ 
from the perturbation
$\delta \theta $ in the derivative  $\theta_{,\mu}$.
Then the second term in (\ref{eq413}) vanishes 
and applying the standard procedure \cite{bilic} 
we can recast (\ref{eq413}) into the form
\begin{equation}
\frac{1}{\sqrt{-G}}
\partial_{\mu}
\left(
{\sqrt{-G}}\,G^{\mu\nu}
 \partial_{\nu}\chi\right) + m_{\rm eff}^2 \chi
=0.
\label{eq5}
\end{equation}
Here, the matrix
\begin{equation}
G^{\mu\nu}=\frac{m^2 c_{\rm s} w}{n}
[g^{\mu\nu}-(1-\frac{1}{c_{\rm s}^2})u^\mu u^\nu] ,
\label{eq2208}
\end{equation}
is the inverse of
the effective metric tensor 
\begin{equation}
G_{\mu\nu}=\frac{n}{m^2 c_{\rm s} w}
[g_{\mu\nu}-(1-c_{\rm s}^2)u_\mu u_\nu]\, ,
\label{eq3008}
\end{equation}
with determinant
\begin{equation}
G\equiv \det G_{\mu\nu} =\frac{n^4}{m^8 w^4 c_{\rm s}^2}\det g_{\mu\nu} .
 \label{eq421}
\end{equation}
The mass parameter $m$
in (\ref{eq5})--(\ref{eq421})
is introduced to make $G_{\mu\nu}$ dimensionless.  
The effective mass squared is given by 
\begin{equation}
m_{\rm eff}^2= m^2\frac{c_s w^2}{n^2}
\left[\left( f
\frac{\partial n}{\partial s}u^\mu\right)_{;\mu}
-f\frac{\partial}{\partial s}\left(f\frac{\partial p}{\partial s}\right)\right] ,
 \label{eq420}
\end{equation}
and the quantity $c_{\rm s}$ is the so-called ``adiabatic'' speed of sound
defined as
\begin{equation}
c_{\rm s}^{2}= \left.\frac{\partial p}{\partial \rho} \right|_s\equiv
\frac{n}{w}\left( \frac{\partial n}{\partial w}\right)^{-1}
=\frac{{\cal L}_X}{{\cal L}_X+2 X{\cal L}_{XX}}. 
\label{eq2011}
\end{equation}
Hence, the linear perturbations $\chi$ propagate in the effective metric 
(\ref{eq3008})
and acquire an effective mass. 
If we replace 
 \begin{equation}
 f\frac{\partial}{\partial s} \rightarrow \frac{\partial}{\partial\theta}
  \label{eq440},
 \end{equation}
 equation (\ref{eq5}) with (\ref{eq2208})-(\ref{eq2011}) will coincide with that of
 Ref.\  \cite{babichev} derived in a different way for a general Lagrangian of the form
 $\mathcal{L}=\mathcal{L}(X,\theta)$.
Note that 
the particle number density $n$ and specific enthalpy $w$
in our notation differ from those of  Ref.\ \cite{babichev} by factors $\sqrt{2}$
and $1/\sqrt{2}$, respectively owing to a factor of $\sqrt{2}$ difference 
in the definition (\ref{eq403}) of the velocity potential. 

\subsection{Purely adiabatic perturbations with $\sigma=0$}
To study the propagation of sound in an inhomogeneous medium
one can consider only adiabatic perturbations and neglect the nonadiabatic ones.
In this case 
by redefining the perturbation $\chi \rightarrow \tilde{\chi}$ so that
\begin{equation}
\chi_{,\mu}=h \tilde{\chi}_{,\mu},
 \label{eq422}
\end{equation}
we will seek the function $h=h(w,s)$ such that
the unwanted term (\ref{eq419})  is eliminated  
from (\ref{eq413}).
Substituting (\ref{eq422}) into (\ref{eq413}) with $\sigma=0$ we find
\begin{equation}
h\left(f^{\mu\nu}
\tilde{\chi}_{,\nu} \right)_{;\mu}
+\left(f^{\mu\nu}h_{,\mu}
-fh\frac{\partial n}{\partial s}u^\nu \right)\tilde{\chi}_{,\nu}
=0.
 \label{eq424}
\end{equation}
Now we demand that the second term in this equation vanishes identically.
Since $\tilde{\chi}_{,\mu}$ is basically arbitrary, this term will vanish if and only if
the function $h$ satisfies
\begin{equation}
f^{\mu\nu}h_{,\mu}
=fh\frac{\partial n}{\partial s}u^\nu .
 \label{eq425}
\end{equation}
Noting that 
\begin{equation}
h_{,\mu}
=\frac{\partial h}{\partial w} w_{,\mu}+\frac{\partial h}{\partial s}s_{,\mu} =
\frac{\partial h}{\partial w} w_{,\mu}+f w\frac{\partial h}{\partial s}u_\mu ,
 \label{eq426}
\end{equation}
where the second equality follows from  (\ref{rnif10}) and (\ref{eq437}),
equation (\ref{eq425}) can be recast into the form
\begin{equation}
\frac{\partial h}{\partial w} f^{\mu\nu} w_{,\mu}+f\left(\frac{\partial h}{\partial s}\frac{n}{c_{\rm s}^2} 
- h\frac{\partial n}{\partial s}\right) u^\nu=0.
 \label{eq427}
\end{equation}
Since the vector $f^{\mu\nu}w_{,\mu}$ is generally not parallel to $u^\nu$, the above identity will 
hold true if and only if the function $h$ does not depend on $w$ and satisfies 
\begin{equation}
 \frac{1}{h}\frac{\partial h}{\partial s}=\frac{c_{\rm s}^2}{n}\frac{\partial n}{\partial s} .
 \label{eq428}
\end{equation}
Clearly, this identity can hold true only if its right-hand side does not depend on $w$.
This together with the definition of the adiabatic speed of sound (\ref{eq2011}) 
yields a condition that the quantity
\begin{equation}
\frac{1}{w}\frac{\partial n/\partial s}{\partial n/\partial w}
 \label{eq434}
\end{equation}
must be  a function of $s$ only.
Applying a very general ansatz
\begin{equation}
n(w,s) =\sum_\alpha w^\alpha \varphi_\alpha(s),
 \label{eq435}
\end{equation}
where  $\alpha$ can be integers or non-integers and $\varphi_\alpha$ are  functions of $s$,
we find that $n$ must be  a function of the form
\begin{equation}
n(w,s) =\sum_\alpha C_\alpha w^\alpha \varphi(s)^\alpha=n(x).
 \label{eq436}
\end{equation}
where $x=w\varphi(s)$, the quantities $C_\alpha$ in the sum are real coefficients, and $n(x)$ and $\varphi(s)$ are 
arbitrary functions
of single variables $x$ and $s$, respectively. 
From (\ref{eq428}) and (\ref{eq2011}) it follows
\begin{equation}
h(s)={\rm const} \,\varphi(s) .
 \label{eq429}
\end{equation}
From (\ref{rnif5}), (\ref{eq436}), and (\ref{eq429})
 we deduce that the pressure must be of the form
\begin{equation}
p=\frac{1}{h(s)} F(w h(s))-V(s) ,
 \label{eq430}
\end{equation}
where $h$ and  $V$ are arbitrary functions of $s$ and $F(x)$ is an 
indefinite integral of $n(x)$, i.e.,
$n=dF/dx$.
The energy density is then fixed by  (\ref{eq404}) 
\begin{equation}
\rho=w n-p.
 \label{eq432}
\end{equation}
Thus, the second term in (\ref{eq424}) can vanish identically if and only if
the pressure is of the form (\ref{eq430}).
Then, we can write (\ref{eq424}) in the form
of a massless Klein-Gordon equation 
\begin{equation}
\frac{1}{\sqrt{-G}}
\partial_{\mu}
\left(
{\sqrt{-G}}\,G^{\mu\nu}
 \partial_{\nu}\tilde{\chi}\right) 
=0
\label{eq50}
\end{equation}
in an effective curved background described by the metric 
(\ref{eq3008}).

In the language of field theory the pressure (\ref{eq430}) corresponds to the Lagrangian
of the form
\begin{equation}
\mathcal{L}=\frac{1}{{h}(\theta)}F({h}(\theta) \sqrt{X}) - {V}(\theta),
 \label{eq433}
\end{equation}
where ${h}(\theta) =h(s(\theta))$, ${V}(\theta)=V(s(\theta))$, and $s(\theta)$
is a function that satisfies (\ref{eq438}).
This form includes 
the trivial case $h={\rm const}$, i.e., $\partial n/\partial \theta=0$,   and
the case of an isentropic fluid:   $V=0$ and $F(x) \propto x^\alpha$. 

\section{Analog planar black hole}
\label{planar}


As an application of  the formalism presented in sections \ref{nonisentropic} and \ref{acoustic},
in this section we address the model of an analog  planar BH hole in asymptotic AdS$_5$
which may have interesting applications 
in condensed matter physics \cite{hartnoll}.
This model  was discussed in detail 
 by S.~Hossenfelder \cite{hossenfelder,hossenfelder2}.
 In her approach, a conservation of particle number is imposed
so the fluid is required to be isentropic. However, in order to maintain the 
energy-momentum conservation, i.e., the Euler equation and a correct 
definition for the speed of sound, it is necessary to 
introduce an external pressure field. This in turn implies a violation of the Poincar\'e invariance 
of the Lagrangian in the field theoretical formulation.

In our approach we will consider a fluid with no external pressure field.
We will  demonstrate  that this model then yields  
a nonisentropic fluid  and derive a Poincar\'e invariant Lagrangian 
that reproduces the desired analog metric.

We start from a planar AdS$_5$ BH with
line element \cite{hartnoll} 
\begin{eqnarray}
{\rm d}s^2 = \frac{\ell^2}{ z^2}  \left[\gamma(z)   dt^2 - 
\gamma( z)^{-1} dz^2 -  \sum_{i=1}^{3} {\rm d} x^i {\rm d} x^i \right],
\label{eq200}
\end{eqnarray}
where $\ell$ is the curvature radius of AdS$_5$,
\begin{eqnarray}
 \gamma(z) =  1 - \left( \frac{ z}{z_0}\right)^4 ,
 \label{gammach}
\end{eqnarray}
and $z_0$ is the location of the  BH horizon.
Following \cite{hossenfelder} we seek a fluid analog model on a 3+1 dimensional
slice perpendicular to the BH horizon  which would mimic the induced metric of the form (\ref{eq200})
with the sum $\sum_{i=1}^{3}$ replaced by $\sum_{i=1}^{2}$.
The basic idea is to find a suitable coordinate transformation
$t\to \tilde{t}$, $z\to \tilde{z}$ such that the new metric 
takes the form of the relativistic acoustic metric (\ref{eq3008})
with $g_{\mu\nu}$ replaced by the Minkowski metric $\eta_{\mu\nu}$.
It has been shown \cite{hossenfelder} that this goal can be achieved by the
 coordinate transformation 
\begin{equation}
 t=\tilde{t}+ f(z), \quad  z = z(\tilde{z}) .
 \label{eq120}
\end{equation}
where the functions $z(\tilde{z})$ and $f(z)$ are determined by the requirement that
the transformed metric takes the form (\ref{eq3008}).
Then,  the speed of sound and the nonvanishing components of the velocity vector
$u_{\tilde{t}}$ and $u_{\tilde{z}}$  in transformed coordinates are given by
\begin{equation}
 c_{\rm s}\equiv dz/d\tilde{z},
 \label{eq107}
\end{equation}
\begin{equation}
u_{\tilde{t}}= \frac{(1-\gamma)^{1/2}}{(1-c_{\rm s}^2)^{1/2}}, \quad 
u_{\tilde{z}}=- \frac{(c_{\rm s}^2-\gamma)^{1/2}}{(1-c_{\rm s}^2)^{1/2}},
 \label{eq123}
\end{equation}
provided that the function $f(z)$  satisfies
\begin{equation}
 \frac{df}{dz}=-\frac{1-c_{\rm s}^2}{\gamma c_{\rm s}}u_{\tilde{t}}u_{\tilde{z}}.
 \label{eq207}
\end{equation}
Next, by applying the potential-flow equation (\ref{eq403})
 we derive closed expressions for $w$, $n$, and $c_{\rm s}$ 
in terms of the variable $z$. Since the metric is stationary, the velocity potential must be of the form
\begin{equation}
 \theta=m\tilde{t} + g(z)
 \label{eq201}
\end{equation}
where $m$ is an arbitrary mass and $g(z)$ is a function of $\tilde{z}$ through $z$.
Then from (\ref{eq403}) it follows
\begin{equation}
 w=\frac{m}{u_{\tilde{t}}}=m\frac{1}{y}(1-c_{\rm s}^2)^{1/2},
 \label{eq101}
\end{equation}
where we have introduced a dimensionless variable
\begin{equation}
y=\frac{z^2}{z_0^2} .
 \label{eq103}
\end{equation}
Besides, it follows from (\ref{eq403}) that the function $g$ in (\ref{eq201}) must satisfy
\begin{equation}
\frac{dg}{dz}=\frac{w}{c_{\rm s}} u_{\tilde{z}}=-\frac{m}{c_{\rm s}y}(c_{\rm s}^2-\gamma)^{1/2}.
 \label{eq121}
\end{equation}
Since the conformal factor in (\ref{eq200}) must be equal to that of (\ref{eq3008}),  i.e.,
\begin{equation}
\frac{n}{m^2 c_{\rm s} w}=\frac{\ell^2}{ z^2} ,
\label{eq202}
\end{equation}
using (\ref{eq101}) one can also express $n$ in terms of $y$ and $c_{\rm s}$.
In this way both $w$ and  $n$  are expressed as functions of $y$ and  $c_{\rm s}$.
However, $c_{\rm s}$ is not independent since by the definition (\ref{eq2011})
\begin{equation}
c_{\rm s}^{2}= 
\left.\frac{n}{w} \frac{\partial w}{\partial n}\right|_s =\frac{n}{w} \frac{w_{,y}}{n_{,y}}.
\label{eq3011}
\end{equation}
where  the subscript $,y$ denotes a derivative with respect to $y$.

At this point we depart from Ref.\ \cite{hossenfelder} in which the continuity equation $(nu^\mu)_{;\mu}=0$
was imposed.
Instead, we require 
a strict validity of (\ref{eq3011}) and thus satisfying the Euler equation without
introducing an external pressure field. As a consequence, the continuity equation 
in our model is not satisfied and the fluid is essentially nonisentropic.

The derivatives of $w$ and $n$ with respect to $y$ may be easily calculated using (\ref{eq101}) and (\ref{eq3008}) 
and using (\ref{eq3011}) a
simple differential equation for $c_{\rm s}$ is obtained 
with solution
\begin{equation}
c_{\rm s}^2=c_1 y^2 +1/2 .
\label{eq106}
\end{equation}
where the integration ``constant'' $c_1$ is generally a function of $s$ 
and must satisfy the restriction $-1/2 \leq c_1\leq 1/2$.
Plugging (\ref{eq106}) into (\ref{eq101}) and (\ref{eq202}) one obtains $w$ and $n$ as functions of $y$
\begin{equation}
w=m\left(\frac{1}{2y^2}-c_1\right)^{1/2} ,
 \label{eq115}
\end{equation}
\begin{equation}
n=m^3\frac{\ell^2}{z_0^2}\left(\frac{1}{4y^4}-c_1^2\right)^{1/2} .
 \label{eq116}
\end{equation}
Now,  one can easily verify   that
\begin{equation}
\partial_{y}(u_{\tilde{z}}n)\neq 0,
 \label{eq124}
\end{equation}
hence, the particle number is not conserved. 
According to Eq.\  (\ref{eq442}), a  non-conservation of the particle number 
automatically implies a nonisentropic fluid.

Note that explicit functional forms of $z(\tilde{z})$, $f(z)$, and $g(z)$ can be obtained
by making use of (\ref{eq106}) and integrating respectively (\ref{eq107}), (\ref{eq207}), and (\ref{eq121}).
However, the precise forms of these functions are not really needed for obtaining the closed expression
for the analog metric.

For the field theoretical description an important quantity is the pressure.
The pressure may be derived from  the equation
\begin{equation}
p_{,y}=nw_{,y},
 \label{eq112}
\end{equation}
which follows from the first thermodynamic relation in (\ref{eq418}).
With the help of (\ref{eq115}) and (\ref{eq116}) one finds
a differential equation 
\begin{equation}
p_{,y}=-\frac{\ell^2}{z_0^2}\frac{m^4}{2y^3}\left(\frac{1}{2y^2}+c_1\right)^{1/2}
 \label{eq113}
\end{equation}
which may be easily integrated.
It is convenient to express $p$ as a function $p=p(w,s)$. Using (\ref{eq113}) and (\ref{eq115}) one finds
\begin{equation}
p=\frac43\frac{\ell^2m^4}{z_0^2}\left(\frac{w^2}{m^2}+2 c_1(s)\right)^{3/2}+c_2(s) .
 \label{eq118}
\end{equation}
where the integration ``constants'' $c_1$ and $c_2$ are arbitrary functions of $s$
(up to the restriction on $c_1$  mentioned above).
Note that with a particular  choice $c_1=0$ the fluid would belong  to the class 
described by (\ref{eq430}) with $h={\rm const}$.

 Next we derive a Lagrangian that reproduces the desired fluid flow and analog metric.
It is quite straightforward to apply the general considerations of section \ref{field} to the model described above.
First, using Eqs. (\ref{eq409}) and (\ref{eq112}) one can determine a functional relationship
between $X$ and  variable $y\equiv z^2/z_0^2$. 
Next, using this relation one can easily find field theoretical representations
of all other fluid functions, such as $c_{\rm s}$, $n$, $p$, and $\rho$.
In particular, the pressure $p$ yields the desired  Lagrangian $\mathcal{L}$  if we replace
$w$ in (\ref{eq118}) by $\sqrt{X}$ and, assuming a functional relationship $s=s(\theta)$,
replace the functions $c_1(s)$
and $c_2(s)$ by functions of $\theta$. In this way, we  obtain
\begin{equation}
\mathcal{L}=\frac43\frac{\ell^2 m^4}{z_0^2}\left(\frac{X}{m^2}+V_1(\theta) \right)^{3/2}+V_2(\theta)
 \label{eq119}
\end{equation}
where we have identified 
\begin{equation}
V_1(\theta)=2c_1(s), \quad V_2(\theta)=c_2(s) .
 \label{eq220}
\end{equation}
Again, with the choice $V_1=0$ this field theory model would belong to the class of models
described by the Lagrangian  (\ref{eq433}) with $h(\theta)={\rm const}$.
Finally, one can easily verify that the acoustic metric (\ref{eq3008}) and hence
the initial planar BH metric (\ref{eq200}) are correctly reproduced by (\ref{eq119})
provided the function $\theta(x)$ defined by (\ref{eq201}) is a solution to the equation of motion
(\ref{eq400}).

Note that the $X$ dependence of the Lagrangian (\ref{eq119}) is the same
as that of the Lagrangian  derived in \cite{hossenfelder}. However,  in contrast to \cite{hossenfelder},
our Lagrangian is Poincar\'e invariant as there is no explicit $z$-coordinate dependence.

\section{Conclusions}
\label{conclude}

We have demonstrated that the formalism of analog gravity
under certain conditions 
can be extended to the case of nonisentropic fluids.
First, if the flow is such that the fluid dynamics can be equivalently described by a scalar field theory
 the acoustic geometry can be fully applied
but a phonon propagating in the fluid generally becomes effectively massive. 
In this case the nonadiabatic perturbations are identified with 
adiabatic ones.
Second, if the nonadiabatic perturbations are neglected,
the standard equations of analog acoustic geometry apply also for a nonisentropic flow 
with  pressure of the form (\ref{eq430}) in which case the phonons remain massless.

As a concrete example, we have applied our nonadiabatic formalism to the analog model of
a planar BH in AdS$_5$.

\appendix
\section{The second law of thermodynamics}
\label{second}

In general the second law of thermodynamics is a global law, which only tells that the entropy 
of the whole system cannot decrease. 
Different subsystems can exchange heat, so the entropy of a subsystem may decrease. 
In an ideal fluid, however, there in no exchange of heat between different 
parts of the fluid \cite{landau}. 
Noting that the entropy density equals $ns$, the second law of thermodynamics in an ideal fluid takes a local form
\begin{equation}\label{rnif12}
(ns u^{\mu})_{;\mu}\geq 0 .
\end{equation}  
Since 
$(ns u^{\mu})_{;\mu}=s(n u^{\mu})_{;\mu}
+nu^{\mu}s_{,\mu}$, Eq.~(\ref{rnif7}) gives
\begin{equation}\label{rnif15}
(ns u^{\mu})_{;\mu}=-\frac{g}{T} (n u^{\mu})_{;\mu} ,
\end{equation}
where
\begin{equation}\label{rnif16}
g=w-Ts
\end{equation}
is the specific Gibbs free energy. Thus the second law (\ref{rnif12}) is equivalent to the condition
\begin{equation}\label{rnif18}
g(n u^{\mu})_{;\mu}\leq 0 .
\end{equation}

To clarify the physical meaning of (\ref{rnif18}), it is instructive to consider the case of a homogeneous 
fluid in the Minkowski background. 
In comoving coordinates, Eq.~(\ref{rnif7}) reads 
$w\partial_{t}n+Tn\partial_{t}s=0$. Assuming that the system has a fixed
volume $V$, we can multiply this equation by $V$ and write it as
\begin{equation}\label{rnif7’’}
wdN+TNds=0 ,
\end{equation}
where $N=Vn$ is the total number of particles in the volume $V$. Eq.~(\ref{rnif7’’}) states that, 
in a system  with conserved energy and positive $w,$ 
the entropy per particle increases when the number of particles decreases. Using $S=Ns$ we have $dS=sdN+Nds$, 
so Eq.~(\ref{rnif7’’}) becomes
\begin{equation}\label{rnif15’}
dS=-\frac{g}{T}dN .
\end{equation}
Therefore the second law of thermodynamics $dS\geq 0$ is equivalent to
\begin{equation}\label{rnif18’}
gdN\leq 0. 
\end{equation}
Eqs.~(\ref{rnif15}) and (\ref{rnif18}) are nothing but 
local covariant versions of (\ref{rnif15’}) and (\ref{rnif18’}), respectively.


Physically, Eq.~(\ref{rnif18’}) can be understood as a result of a competition between two effects. 
First, for a fixed entropy per particle the total number of particles increases with increasing total entropy.
Second, owing to (\ref{rnif7’’}), the total number of particles decreases with increasing entropy per particle.
Thus, depending on which effect prevails, the number of particles will increase or decrease.
Equations (\ref{rnif18’}) or (\ref{rnif18}) tell us that  the sign of the Gibbs free energy is crucial: 
the number of particles will not increase (decrease)
if the Gibbs free energy is positive (negative).

\section*{Acknowledgments}

This work has been supported by
the H2020 CSA Twinning project No.\ 692194, ``RBI-T-WINNING''.
The work of N.B.\ has been partially supported by the ICTP - SEENET-MTP project NT-03 
Cosmology - Classical and Quantum Challenges.

\end{document}